\documentclass[preprint,prl]{revtex4}

\usepackage{graphicx}

\newcommand{\figurewidth}{0.5\textwidth}

\begin{document}

\title{Unfolding Collapsed Polyelectrolytes in Alternating-Current Electric Fields}
\author{Pai-Yi Hsiao}
\email[Corresponding author, E-mail: ]{pyhsiao@ess.nthu.edu.tw}
\author{Yu-Fu Wei}
\affiliation{%
    Department of Engineering and System Science, 
    National Tsing Hua University, 
    Hsinchu, Taiwan 30013, R.O.C.
}
\author{Hsueh-Chia Chang}
\email[E-mail: ]{hchang@nd.edu}
\affiliation{%
    Department of Chemical and Biomolecular Engineering, 
    University of Notre Dame, 
    IN 46556, U.S.A.
}

\date{\today}

\begin{abstract}
We investigate the unfolding of single polyelectrolyte (PE) chains collapsed by
trivalent salt under the action of alternating-current (AC) electric fields
through computer simulations and theoretical scaling. The results show that a
collapsed chain can be unfolded by an AC field when the field strength exceeds
the direct-current (DC) threshold and the frequency is below a critical value,
corresponding to the inverse charge relaxation/dissociation time of condensed
trivalent counterions at the interface of the collapsed electrolyte.  This
relaxation time is also shown to be identical to the DC chain fluctuation time,
suggesting that the dissociation of condensed polyvalent counterion on the
collapsed PE interface controls the polyelectrolyte dipole formation and
unfolding dynamics under an AC electric field .
\end{abstract}

\maketitle 

\section{I. Introduction}
To develop new ability to separate and stretch biomolecules, such as nucleic
acids and proteins, is very important in molecular biology and medical research
for its applications in single molecule analysis and DNA
sequencing~\cite{shendure04,chan05,gupta08,treffer10}.  Conventional way to
determine the genetic information of a DNA molecule relies on a series of
cumbersome operations of subcloning and electrophoretic
separation~\cite{sanger77}. This current method reads the sequence of DNA
fragments of hundred base-pairs per time, which has become inadequate for
individualized genetic studies~\cite{chan05}.  To improve the efficiency of
sequencing, single-molecule sequencing methods have been developed
recently~\cite{chan05,gupta08,treffer10}, in which the genetic information of
individual DNA molecules is scanned in a linear fashion. One of the key
problems in these methods concerns the unfolding of DNA molecules to increase
the spatial resolution of scanning, as DNA molecules are long polyelectrolyte
(PE) chains that usually adopt the coiled conformation in a typical buffer.  In
the past decades, many techniques have been developed to stretch PE chains,
including the utilization of optical tweezers~\cite{smith96}, fluid
flows~\cite{perkins95}, electric
fields~\cite{washizu90,washizu95,ferree03,kaji02,kaji03}, and other
methods~\cite{tegenfeldt01,bensimon94,smith92}.  Among these methods,
stretching PE chains in electric fields has received considerable attention, as
electrokinetics has become a viable microfluidic method for on-chip molecular
analysis~\cite{hughes00,chang10}.

DNA electrophoresis is usually performed in a sieving medium, such as a gel or
a polymer solution~\cite{cottet98,viovy00}. In that case, DNA chains transform
cyclically between an extended structure, with one end hooked onto obstacles in
the medium, and a coiled structure when released from the
obstacles~\cite{smith89}. If an alternating-current (AC) electric field of
appropriate frequency is applied instead, the conformation of chain in a
sieving medium can be maintained constantly in an elongated
structure~\cite{kaji02, kaji03}. In a bulk buffer, coiled DNA chains have
been demonstrated to be stretched in a strong direct-current (DC) electric
field, through the mechanism of polarization~\cite{porschke85b,washizu90}.
Nevertheless, there is a drawback to this method: a strong DC field leads
easily to electrolytic dissociation of water molecules~\cite{washizu90,
chang07}. To avoid this drawback, a short-duration pulse or an AC electric
field has been suggested, with the period shorter than the dissociation
reaction time. Moreover, the mean displacement of PE chains in an AC field is
zero. This feature could be used to trap or to manipulate single DNA molecules
at some place to facilitate the detection of the sequence, particularly when it
is integrated with a dielectrophoresis trapping mechanism~\cite{chang10}.
Experiments have shown that AC electric field can indeed stretch and orient DNA
chains in aqueous solutions at low AC frequency and the chain polarization is
largely reduced while the frequency is elevated~\cite{washizu04,dukkipati07}.
Recently, Wang et al.~demonstrated that the conformational transition of single
PEs possesses a hysteretic nature upon sweeping the frequency of AC electric
field~\cite{wang10}.    

Without a solid medium to break the symmetry of the coil-stretch transitions, a
low-frequency AC field should stretch the DNA repeatedly during every
half-cycle, in a manner of a quasi-DC field. However, unlike DC stretching, the
polarization and stretching dynamics now becomes important, as the molecule
must be fully stretched within one half period.  It is known that multivalent
cations have dramatic effects on the properties and structures of DNA molecules
in aqueous solutions~\cite{bloomfield96}. Coiled DNA chains can collapse into
globule particles while the chain charge is neutralized by the condensed
multivalent cations~\cite{bloomfield96,hsiao06c}. Experiments have shown that a
collapsed DNA chain can be unfolded in DC electric fields if the field strength
surpasses a threshold~\cite{porschke85a}. This threshold $E^*$ depends on the
chain length $N$, and the dependence has been studied recently by computer
simulations~\cite{netz03a,netz03b,hsiao08,wei09}.  The results suggested a
power law-like relation $E^* \sim N^{-\theta}$ and the electrophoretic mobility
of chain depends strongly on whether the chain is unfolded or not.  These
studies provide a solid foundation to separate PE chains by length in free
solutions, based upon the drastic mobility change when a chain is unfolded by
an applied electric field~\cite{hsiao08}.  Moreover, recent study in protein
mass spectrometry showed that AC electrospray ionization (ESI) yields an
effective ionization without fragmenting molecules, with higher ionization
efficiency than DC ESI~\cite{chetwani10}. The ionization efficiency is a
sensitive function of sample pH and molecular conformation. It is still unclear
whether the higher efficiency is due to the increased local pH or the
conformational change or both of the two effects.  

This present work constitutes a fundamental molecular dynamics study of how an
AC field affects the conformation of a PE chain and the condensation of
counterions, and how the field strength to unfold the chain varies with
frequency, in order to give deep insight of the complicated behavior of PE
solutions in AC fields. The rest of the paper is organized as follows.  We
describe the model and simulation setup in Sec.~II.  The results and
discussions are presented in Sec.~III.  The topics include: chain unfolding in
AC electric fields (Sec.~III.A), simulation snapshots (Sec.~III.B), theoretical
explanation (Sec.~III.C), time evolution of dipole moment and chain unfolding
(Sec.~III.D), and the study of polarization time and fluctuation time
(Sec.~III.E).  A summary is given in Sec.~IV.  

\section{II. Simulation model and setup}
Our system contains a single linear chain, modeled by a bead-spring chain
model.  The chain is composed of $N$ monomers. Each monomer carries a negative
unit charge $-e$ and dissociates one monovalent counterion into the solution.
The bond connection on the chain is modeled by the finitely extensible
nonlinear elastic potential
\begin{equation}
U_{\rm FENE}(b)=-\frac{1}{2}k b_{max}^2\ln\left(1-\frac{b^2}{b_{max}^2}\right)
\end{equation}
where $b$ is the bond length between two neighboring monomers, $b_{max}$ the
maximum extension, and $k$ the spring constant.  The condensation of the PE
chain is induced by adding (3:1)-salt into the system.  The salt molecules
dissociate into trivalent cations (also called ``counterions'') and monovalent
anions (also called ``coions'') in the solution.  All these particles,
including monomers, counterions, and coions, are modeled as Lennard-Jones (LJ)
spheres, described by the potential
\begin{equation}
U_{\rm LJ}(r)= \left\{ \begin{array}{ll} 4\varepsilon_{\rm LJ}\left[
\left(\sigma/r\right)^{12} -\left(\sigma/r\right)^{6} + 1/4 \right] & \mbox{\ ,
for $r \le \sqrt[6]{2} \sigma \; $}\\ 0 & \mbox{\ , for $r > \sqrt[6]{2} \sigma
\; $} \end{array} \right.
\end{equation}
where $\sigma$ and $\varepsilon_{\rm LJ}$ represent the diameter and the
hardness of the LJ sphere, respectively, and $r$ is the distance between two
particles.  Since the interaction between the monomers is purely repulsive,
this setup corresponds to a good solvent condition.  Particles also interact
with each other via Coulomb interaction, which read as
\begin{equation}
U_{\rm C}(r)= k_B T \cdot \frac{Z_i Z_j\lambda_B}{r}
\end{equation}
where $k_{B} T$ is the thermal energy and $Z_i$ is the charge valence of the
$i$th particle.  The Bjerrum length $\lambda_B = e^2/(4\pi \epsilon_0
\epsilon_r k_B T)$ is the distance between two unit charges, at which the
electrostatic energy is equal to the thermal energy, where $\epsilon_0$ is the
vacuum permittivity and $\epsilon_r$ is the relative dielectric constant of
solvent. The solvent molecules are not modeled explicitly in the study.
However, their effects are incorporated implicitly by the following three ways:
(1) dielectric screening of charge by $\epsilon_r$, (2) friction force,
$-m_i\zeta_i\vec{v}_i$, due to particle moving through the solvent, (3)
stochastic force, $\vec{\eta}_i(t)$, owing to thermal collisions of solvent
molecules.  The equation of motion, also known as the Langevin equation, thus
reads as
\begin{equation}
m_i \ddot{\vec{r}}_i = -m_i\zeta_i\dot{\vec{r}}_i +\vec{F}_c +Z_i e E(t) {\hat
x} +\vec{{\eta}}_i \label{eq:langevin}
\end{equation}
where $m_i$ is the particle mass, $m_i\zeta_i$ is the friction coefficient, and
$\vec{F}_c=-\partial\,U/\partial\,\vec{r}_i$ is the conservative force. In this
equation, the temperature $T$ is controlled by the fluctuation-dissipation
theorem: $\left< \vec{\eta}_i(t) \cdot \vec{\eta}_j(t') \right> = 6 k_B T
m_i\zeta_i\delta_{ij} \delta(t-t')$.  The system is placed in a periodic
rectangular box. The AC field $E(t)$ is a square wave applied in $x$-direction.
The period of the square wave is $\Psi$ and the field strength is $|E|$.
Particle-particle particle-mesh Ewald sum~\cite{hockney88} is used to calculate
Coulomb interaction.

We simulate aqueous PE solutions at room temperature ($T=300$K) and assume that
all the particles have the same mass $m$ and LJ parameters $\sigma$ and
$\varepsilon_{\rm LJ}$. We set $\varepsilon_{LJ}= 0.8333 k_B T$, $k =5.8333
k_BT/\sigma^2$, $b_{max} =2\sigma$, $\lambda_B =3\sigma$, and $\zeta_i =1.0
\tau^{-1}$ where $\tau =\sigma \sqrt{m/(k_BT)}$ is the time unit.  Since
$\lambda_B$ is 7.1 \AA\ for water, the length unit $\sigma$ corresponds to
$2.4$ \AA. The time unit $\tau$ is about 1.5 ps if $m$ is taken to a typical
monomer mass of the order 100 g/mol. This set of parameters can be used to model
flexible PE chains in water, such as sodium polystyrene sulfonate. The
dimension of our simulation box is $153.6\sigma \times 79.06\sigma \times
79.06\sigma$ and the chain length $N$ is $96$. It yields the monomer
concentration $C_m$ equal to $0.0001\sigma^{-3}$, which is about 12 mM in real
unit.  This $C_m$ describes a dilute PE solution. The salt concentration $C_s$
is set to the equivalence point $C_s^*=C_m/3$. It has been shown that at
$C_s=C_s^*$, the trivalent counterions form complexes with the chain, largely
neutralizing the bare chain charge; the chain collapses into a globule
structure~\cite{bloomfield96, hsiao06a, hsiao06b, hsiao06c}. $\Psi$ and $|E|$
are varied over a wide range of value to study their effects on the properties
of the PE system. Langevin equation is integrated by Verlet algorithm with a
time step $\Delta t$ equal to $0.005\tau$~\cite{note-lammps}.  A pre-run phase
takes about $10^7$ time steps to bring the system into a stationary state for each
($\Psi$, $|E|$)-case, followed by a production-run phase of $10^8$ time steps
to cumulate data for analysis.  The hydrodynamic interaction is not included in
this work because it is largely screened out under typical electrophoretic
conditions~\cite{long96,viovy00,tanaka}. Recent study clearly demonstrated this
point~\cite{grass09}.  To simplify the notation, physical quantities in the
following text will be reported in ($\sigma$, $m$, $k_B T$, $e$)-unit system if
not specially mentioned. For e.g., the strength of electric field will be 
described in unit $k_BT/(e \sigma)$.   

\section{III. Results and discussions}
\subsection{A. Degree of chain unfolding}
\begin{figure}[htbp]
\begin{center}
\includegraphics[width=\figurewidth,angle=270]{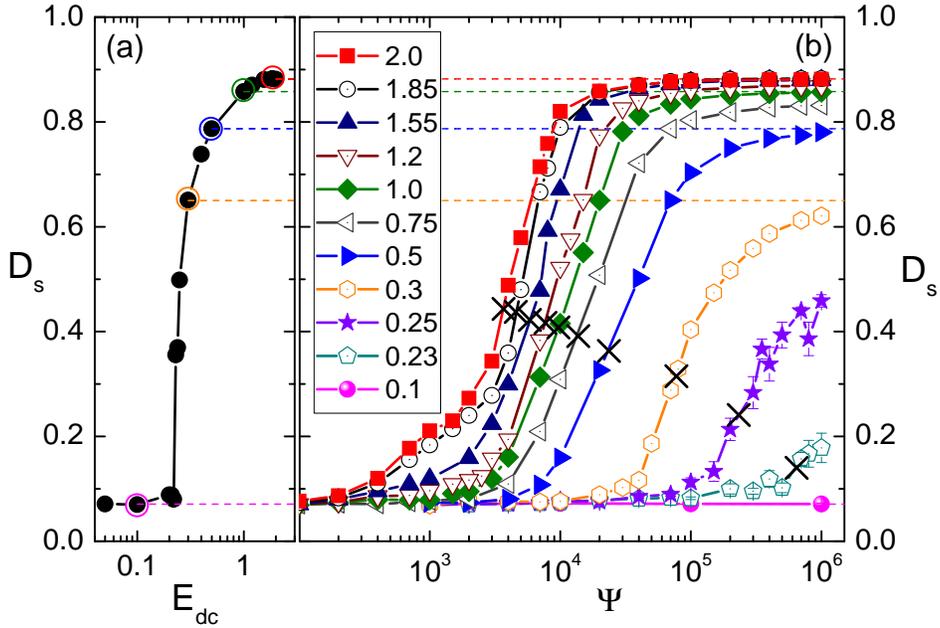}
\caption{(a) $D_s=R_e/L_c$ in DC fields as a function of $E_{\rm dc}$; (b)
$D_s$ as a function of ~$\Psi$ (in unit $\Delta t$) in AC fields of different
$|E|$.  The value of $|E|$ is indicated in the inset of the figure. The symbol
`X' denotes the inflection point on each corresponding curve.  The error bar of
data, if not shown, is smaller than the symbol size of data in this paper.  }
\label{fig:P1N96_Z3_ET_Ds_inflex.eps} \end{center}
\end{figure}

We first revisited a reference case in which the applied field is a DC electric
field. It is the limiting case of an AC field with infinite $\Psi$.  The degree
of chain unfolding $D_s$, defined as the ratio of the end-to-end distance
$R_{e}$ of chain to the contour length $L_c$, is shown in
Fig.~\ref{fig:P1N96_Z3_ET_Ds_inflex.eps}(a) against  the DC field strength
$E_{\rm dc}$.

We can see that $D_s$ shows a sharp transition, occurred at $E_{\rm
dc}^*=0.22$. Below $E_{\rm dc}^*$, the PE chain stays in a collapsed state, in
which condensed trivalent counterions tightly bind monomers together, resulting
in a chain size similar to the unperturbed size in zero field.  The structure
of the collapsed chain is a spherical globule.  When $E_{\rm dc}>E_{\rm dc}^*$,
the chain unfolds abruptly.  The DC field is strong enough to drive the
condensed trivalent counterions towards one side of the chain globule and even
more, strips them off the chain. The chain is hence elongated and aligned in
the field direction in the process. The degree of chain unfolding can be as large as
$90\%$ of the contour length when $E_{\rm dc}$ is large enough. The detailed
study including the effect of salt concentration and valence on chain unfolding
as well as the scaling law for $E_{\rm dc}^*$ against $N$ in DC fields can be
found in Refs.~\cite{hsiao08,wei09}. An  independent simulation done recently
by other group~\cite{liu10} also confirms our results.   
   
We then studied the case with AC fields. The degree of chain unfolding is now
plotted against the field period $\Psi$ in
Fig.~\ref{fig:P1N96_Z3_ET_Ds_inflex.eps}(b) for different field strength $|E|$.
The results show that $D_s$ takes the value of zero field, $0.07$, over the
whole range of $\Psi$ when $|E|<E_{\rm dc}^*$. In this situation, the field
strength is too weak to overcome the binding of the condensed trivalent
counterions to unfold the chain.  When $|E|>E_{\rm dc}^*$, chain unfolding can
take place and $D_s$ is a sigmoidal function of $\Psi$.  When $\Psi$ is small,
the applied field alters very frequently.  The duration of a half period of the
field, within which the square-wave field acts as a DC one pointing to either
$+x$ or $-x$ direction, is not long enough either for the chain to complete its
conformational change or for the condensed counterions to migrate to stable
positions.  The degree of chain unfolding is thus small.  When $\Psi$ is
sufficiently large, the deformation and the polarization of the PE complex can
be completed in a half period. The chain is unfolded and the maximal degree of
chain unfolding takes a value corresponding to the limiting value in the
referenced DC field.  In the figure, the limiting value has been indicated in
dashed line.  The reduction of the degree of chain unfolding in the region of
small $\Psi$ is in agreement with experiments~\cite{washizu04, dukkipati07};
recent simulation~\cite{liu10} also reported a consistent behavior.  Moreover,
we observed that the transition of $D_s$ is not really sharp but occurs over a
range of period $\Psi$. In order to characterize this behavior quantitatively,
we define the transition point at $\Psi=\Psi_c$, which is the inflection point
of the $D_s$ curve. The inflection point has been indicated by a symbol `X' on
each curve in Fig.~\ref{fig:P1N96_Z3_ET_Ds_inflex.eps}(b). We saw immediately
that $\Psi_c$ depends strongly on $|E|$ and the value falls in a range from $3
\times 10^3 \Delta t$ to $7 \times 10^5 \Delta t$, or equivalently, from 22.5
ps to 5.25 ns. Therefore, the collapsed chain is unfolded when the AC field
frequency is smaller than the order of tera- or giga-hertz.  Detailed study
showing that $\Psi_c$ is directly related to the characteristic time to fully
polarize the PE chain will be presented later.  Since chain unfolding occurs in
a small frequency region, experiments can verify this phenomenon easily,
provided that the AC field strength is stronger than $E_{\rm dc}^*$. As having
been shown in our previous DC simulations~\cite{hsiao08,wei09}, $E_{\rm dc}^*$
can be largely reduced to a value of few hundreds volt per cm if a long PE
chain of order, for e.g., $N=10^6$, is used.  This strength of AC field is
accessible by experiments.

\subsection{B. Simulation Snapshots}
\begin{figure}[htbp]
\begin{center}
\includegraphics[width=\figurewidth,angle=270]{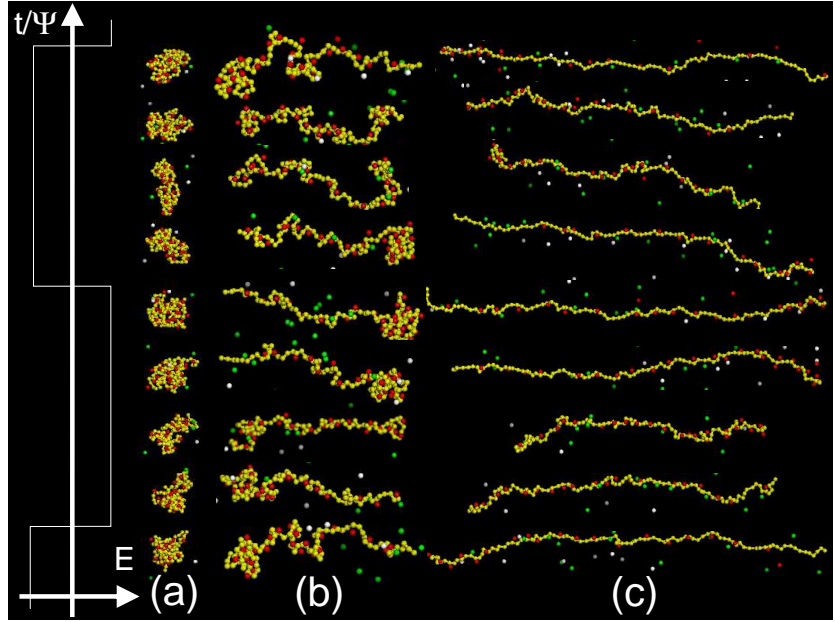}
\caption{(Color online) Snapshots of PE chain in a cycle of the applied AC
field.  The field strength $|E|$ is 0.3 and the field period is (a)
$\Psi=10^{3}\Delta t$, (b) $\Psi=8\times 10^{4}\Delta t$, and (c)
$\Psi=10^{6}\Delta t$. The PE chain is represented by the yellow-colored
bead-spring chain.  The trivalent counterions, the monovalent counterions, and
the coions are represented by the red, green and white beads, respectively.
The change of the field direction is indicated on the left side of the figure
as a function of field cycle $t/\Psi$ and the evolution of the chain
conformation can be seen in each panel, following the $t/\Psi$-axis from bottom
to top. } \label{fig:snapshots} \end{center}
\end{figure}

In order to have the picture of chain conformation in mind, we show in
Fig.~\ref{fig:snapshots} the snapshots of PE chain in AC fields with the field
strength equal to 0.3.

Three field periods are presented in the panels: (a) $\Psi=10^{3}\Delta t$, (b)
$\Psi=8\times 10^{4}\Delta t$, and (c) $\Psi=10^{6}\Delta t$, which are,
respectively, smaller than, about equal to, and larger than the critical period
$\Psi_c$ (refer to Fig.~\ref{fig:P1N96_Z3_ET_Ds_inflex.eps}).  In each panel,
the evolution of chain shape against a field cycle is shown, following the
$t/\Psi$-axis. We can see that for the case of small period
(Fig.~\ref{fig:snapshots}(a)), the chain is not unfolded in the AC field. It
exhibits a globule structure.  This is because the acting time in a half-cycle
of the AC field is too short for the chain to response. On the other hand, for
the case of large period as shown in Fig.~\ref{fig:snapshots}(c), the acting
time is long enough.  The chain is stretched and displays an elongated
structure aligned parallel to the field direction. If we look carefully into
the snapshots, we can find that the trivalent counterions are not uniformly
condensed on the chain backbone in this case. They follow the field direction
and accumulates more on one of the chain end.  This non-uniform distribution
results in a chain polarization.  When the AC field switches its direction, the
stability of the system is suddenly destroyed and the condensed trivalent ions
migrate toward the other chain end. The chain partially shrinks and then
regains its size soon after the new stability is established.  For the case
with the field period close to $\Psi_c$ (Panel (b) of
Fig.~\ref{fig:snapshots}), an astonishing, tadpole-like structure of chain is
observed, just before switching of the AC field direction.  The head of the
tadpole structure is derived from the accumulated trivalent counterions binding
together with the chain near an end.  And the tail is formed by the other side
of the chain with less counterions condensed on it.  Because the chain is
partially unfolded, the degree of polarization is smaller than the one in Panel
(c). The change of the AC field direction drives the trivalent counterions
moving away from the tadpole head toward the tail. As a passage, the head
shrinks and the tail grows.  Eventually, the new accumulation of the trivalent
counterions creates a new head on the other side of the chain. The new tadpole
structure thus inverts its heading direction and waits for the next field
alternating.  

\subsection{C. Theory of chain unfolding}
The dependence of $\Psi_c$ on $|E|$ can be captured by the following charge
relaxation theory.  We assume that a PE chain is unfolded in an AC field if the
polarization energy of the chain complex, $W_{\rm ac}=\alpha(\omega)|E|^2/2$,
is larger than some critical energy, where $\omega$ is the field frequency and
$\alpha(\omega)$ is the polarizability. This critical energy can be estimated
by $W_{\rm dc}^*=\alpha(0)E_{\rm dc}^{*2}/2$, the polarization energy to unfold
the chain in the critical DC field $E_{\rm dc}^*$.  It is known that for a
conducting Maxwell-Wagner sphere of arbitrary radius $R$ in an AC field, the
polarizability can be expressed by $\alpha(\omega)=4\pi\varepsilon_m K(\omega)
R^3$ where $$K(\omega)=
\frac{\tilde{\varepsilon}_p(\omega)-\tilde{\varepsilon}_m(\omega)}
{\tilde{\varepsilon}_p(\omega)+2\tilde{\varepsilon}_m(\omega)}$$ is the
Clausius-Mossotti factor, and $\tilde{\varepsilon}_m$ and
$\tilde{\varepsilon}_p$ are the $\omega$-dependent complex permittivity of the
solvent medium and the PE complex, respectively~\cite{chia_book}.  This
Clausius-Mossotti factor arises due to both the dielectric polarization,
resulting from the permittivity difference across the sphere-medium interface,
and the space charge accumulation at the interface, derived from the
conductivity gradient and the corresponding ion flux discontinuity at that
location.  While counterion migration occurs across the entire collapsed chain,
charge polarization occurs only at the two poles. It involves the dissociation
and condensation of the migrating counterions presented within a Debye-layer
neighborhood of the interface. As a result, the dipole formation time is the
usual charge relaxation time at the interface. When the resulting dipole moment
aligns parallel to the electric field, a net stretching force is imposed on the
chain, corresponding to the polarization energy.  The chain is unfolded.  The
condensed ions hence play a role to the PE chain, similar to the one of the
bound surface electrons to a dielectric object.  Therefore, the polarization
dynamics of the system can be studied by the method of electrical
impedance~\cite{chia_book,basuray07}.  

We model the dielectric response for the PE complex by a resistor-capacitor
(RC) circuit in series. The complex permittivity reads as
$$\tilde{\varepsilon}_p(\omega)
=\frac{\varepsilon_p}{1+i\omega\varepsilon_{p}/\sigma_{p}}
=\frac{\varepsilon_p}{1+i\omega\tau_{rc,p}}.$$ The response function for the
solvent is modeled by a RC circuit in parallel and the complex permittivity is
$$\tilde{\varepsilon}_m(\omega) = \varepsilon_m - \frac{i\sigma_{m}}{\omega}
=\varepsilon_m\left(1-\frac{i}{\omega\tau_{rc,m}}\right).$$ Here
$\varepsilon_{p}$ and $\sigma_{p}$ (or $\varepsilon_{m}$ and $\sigma_{m}$) are,
respectively, the permittivity and the ion conductivity in the PE complex (or
in the solvent), and $\tau_{rc,p}=\varepsilon_{p}/\sigma_{p}$ and
$\tau_{rc,m}=\varepsilon_{m}/\sigma_{m}$. Since the trivalent counterions
condensed quite tightly on the chain, the migration of these ions from one side
of the PE complex to the other side in an electric field is similar to the
electron migration from one plate of a capacitor to the other plate in a
voltage source.  The ion migration also suffers a resistance force against the
applied electric field, a situation which is similar to put a resistor in the
circuit.  Therefore, we model the PE chain phase by an RC circuit in series.
Because the length scale across the phase in the field direction corresponds to
both the capacitor gap and the resistor length, the ratio $\tau_{rc,p}$ is the
characteristic time to move the counterions from one side of the PE chain to
the other side. It corresponds to the RC time of the circuit. In the solvent
phase, the ions travel through the whole system. Since the traveling speed is
not fast, the medium is a poor conductor for ions. Therefore, we regard the solvent
as a leaky dielectric, modeled by an RC circuit in parallel.  A similar argument
shows that the ratio $\tau_{rc,m}$ is the RC time for the solvent.

Due to the linear dependence of the conductivity on the ionic strength, the RC
time for solvent is also the charge relaxation time of electrolyte, which is
equal to the squared Debye screening length divided by the diffusivity of the
dominant ions. For the PE phase, this connection is typically not true because
the ions conduct through nanoporous structures formed by the twisted chain in the
phase, rather than simply through a bulk solution. The RC time for the PE phase
is important in determination of the critical frequency. At high frequency, the
chain polarization is due to dielectric polarization and the value is small.
At low frequency, the ``capacitor'' is fully charged through conductive
polarization by accumulating counterions at one pole of the PE globule,
producing a large polarization. This polarization can grow even further via
chain unfolding if the applied electric field is strong. 

\begin{figure}[htbp]
\begin{center}
\includegraphics[width=\figurewidth,angle=270]{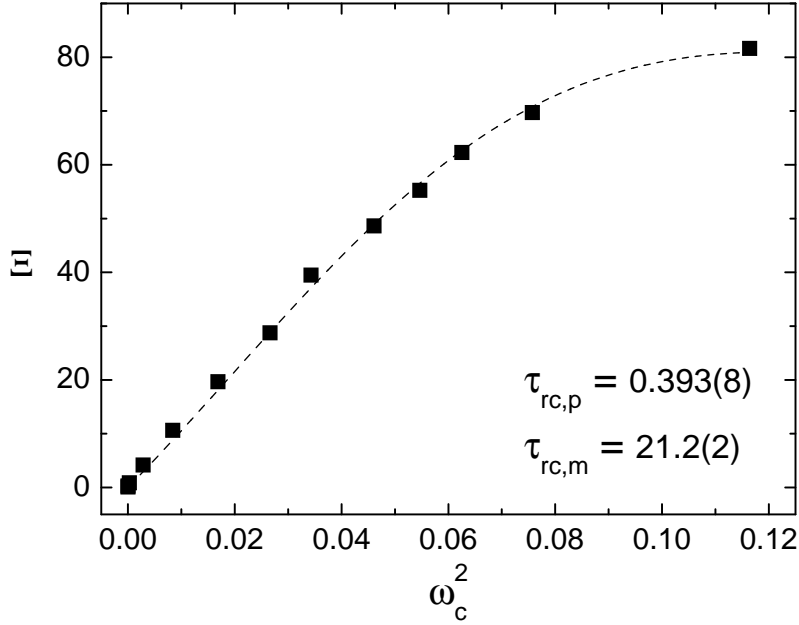}
\caption{$\Xi\equiv (|E|/E_{\rm dc}^*)^2-1$ as a function of the square of the
critical frequency, $\omega_c^{2}$. The dashed curve is the result of fitting
by Eq.~(\ref{eq:Xi}). } \label{fig:P1N96_Z3_w2_Xi_inflex.eps} \end{center}
\end{figure}

We assume that the RC time for PE is a constant at the critical points to
unfold the chain.  Based upon the above assumption for the polarization energy,
the critical frequency $\omega_c$ to unfold a chain is determined by the
criterion $(|E|/E_{\rm dc}^*)^2={\rm Re}[K(0)]/{\rm Re}[K(\omega_c)]$.  Let us
consider the case of $\varepsilon_p \simeq \varepsilon_m$. We obtain
\begin{eqnarray}
\Xi=\frac{(6\tau_{rc,p}\tau_{rc,m}+9\tau_{rc,m}^2)\omega_c^2}
{4+(4\tau_{rc,p}^2-2\tau_{rc,p}\tau_{rc,m})\omega_c^2
+4\tau_{rc,p}^2\tau_{rc,m}^2\omega_c^4} \label{eq:Xi}
\end{eqnarray}
where $\Xi=(|E|/E_{\rm dc}^*)^2-1$ is the excess of the squared electric field.
When $\omega_c$ is small, the equation gives 
$\Xi\simeq\left(\frac{3}{2}\tau_{rc,p}\tau_{rc,m}
+\frac{9}{4}\tau_{rc,m}^2\right)\omega_c^2 +O(\omega_c^{4})$, which is a linear
function of $\omega_c^{2}$.  The figure $\Xi$ vs.~$\omega_c^{2}$ is plotted in
Fig.~\ref{fig:P1N96_Z3_w2_Xi_inflex.eps} where $\omega_{c}$ is calculated by
$2\pi/\Psi_{c}$.

The result does show a linear dependence of $\Xi$ in the region of small
$\omega_c^{2}$, which follows the prediction of our theory.  Least square fit
by Eq.~(\ref{eq:Xi}) using Levenberg-Marquardt algorithm yields
$\tau_{rc,p}=0.393(8)$ and $\tau_{rc,m}=21.2(2)$. In our simulations, the
length unit $\sigma$ and time unit $\tau$ are 2.4\AA\ and 1.5ps, respectively,
while mapped to a real system.  Since the trivalent cations condense on the
chain and largely neutralize the chain charge, most of the monovalent cations
and anions are presented in the bulk solution. Hence, the concentrations of the
cations and the anions in the solution are both about 0.0001 ($\sigma^{-3}$),
or equivalently, 12 mM. The permittivity of water in our system is
$$\epsilon_0\epsilon_r = \frac{e^2}{4\pi k_B T \lambda_B} = \frac{1}{12\pi}
\cdot \frac{e^2}{k_B T\sigma}.$$ Using this value of $\epsilon_0\epsilon_r$
($1/12\pi$ in our unit system) for $\varepsilon_{m}$, the theoretical
interpretation of our coarse-grained simulation of chain stretching yields the
solvent conductivity $\sigma_{m}$ ($=\varepsilon_{m}/\tau_{rc,m})$ equal to
$3.5\times 10^{-5}$, or equivalently, 0.59 S/m~\cite{note-epsilon}.  This
solvent conductivity agrees fairly with the one of sodium chloride in
water~\cite{wu95}, providing a self-consistent support for our theory.  Here we
have corrected the conductivity by dividing the value by a factor 35.  It is
because the friction coefficient in Eq.~(\ref{eq:langevin}) has been set to a
small value $\zeta_i=1.0 \tau^{-1}$ in our study, which is about 35 times
smaller than in water, for the purpose to accelerate the simulations.
Therefore, to map diffusion-related quantities such as conductivity and
relaxation time, we have to correct the obtained values by the factor 35. 

The typical RC relaxation time in experiments ranges from milliseconds to
nanoseconds, depending on the ionic strength~\cite{chia_book,wang10}.  The
ionic strength in our bulk solution is about $I_s=0.5\left((+1)^2\times
10^{-4}+(-1)^2\times 10^{-4}\right)=10^{-4}$ ($\sigma^{-3}$), which gives a
Debye screening length equal to $\ell_D=(4\pi\lambda_B \cdot 2 I_s)^{-1/2}
\simeq 2.76$ nm. Using the typical value of ion diffusivity in water, $D_i \sim
10^{-9}\rm{m^2/s}$~\cite{campbell69}, we expect the relaxation time in water
with a value of about $\ell_D^2/D_i \sim 7.6$ ns. Inside the PE globule, the
ions diffuse through the nanoporous structure formed by the twisted chain. The
size of the nanopores is about $\sigma$, the diameter of the ions.  Therefore,
the relaxation time in PE phase is estimated by $\sigma^2/D_i$ and the value is
58 ps. The RC time, $\tau_{rc,p}$ and $\tau_{rc,m}$ obtained from
Fig.~\ref{fig:P1N96_Z3_w2_Xi_inflex.eps}, is 20.6 ps and 1.1 ns, respectively,
after multiplying the correction factor. These values are in fair accordance
with the theoretical estimation.  We remark that the RC time of
highly-concentrated nanocolloid or molecule suspensions can be measured from
dielectric relaxation spectroscopy or impedance spectroscopy, and the value for
a single nanocolloid or molecule is inferred by an averaging
theory~\cite{chia_book}. Recently, cross-over frequencies of dielectrophoresis
have been used to estimate the RC time of individual blood cells and
nanocolloids~\cite{gordon07, basuray07}. A very recent study has demonstrated 
an approach to determine the RC time of short PE using fluorescent correlation
spectroscopy, based upon the coil-globule transition in AC
fields~\cite{wang10}. 

\subsection{D. Time variation of dipole moment and chain unfolding}
\begin{figure}[htbp]
\begin{center} \includegraphics[width=\figurewidth,angle=270]{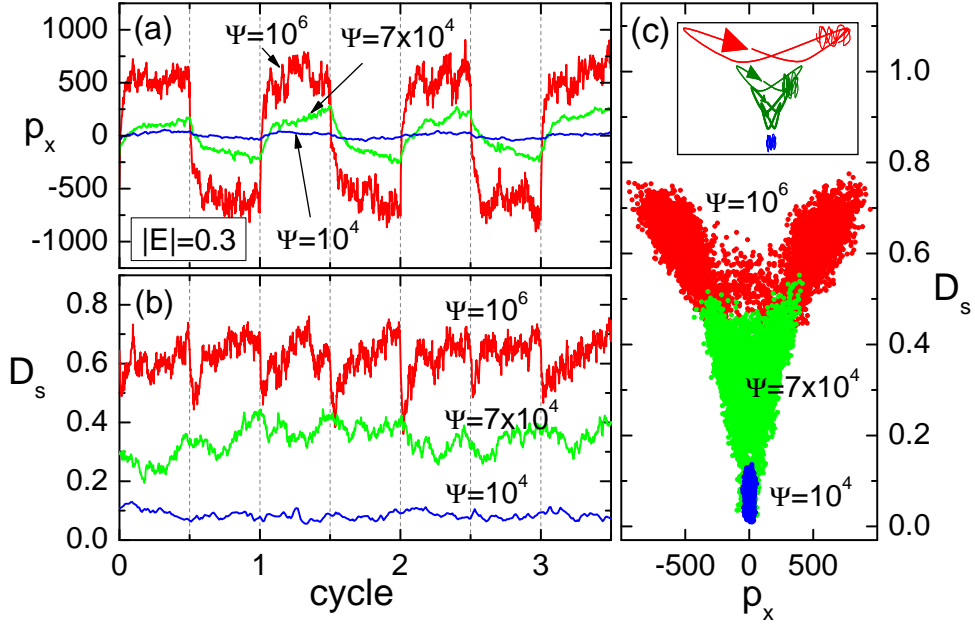}
\caption{Time variation of (a) $p_x$ and  (b) $D_s$, against the cycle of AC
field. (c) Parametric plot of $p_x$ and $D_s$ with time.  The field strength is
$|E|=0.3$. The value of $\Psi$ (in unit $\Delta t$) is indicated near the
associated curve and data. } \label{fig:px-Ds_cycle.eps} \end{center}
\end{figure}

The critical frequency $\omega_c$ obtained in Sec.~III.C is independent of the
radius of gyration of the chain, as the Maxwell-Wagner relaxation time
describes the relaxation/dissociation of ions within a Debye layer from the
interface. This physical picture breakdowns when the chain begins to uncoil,
such that the trivalent counterion no longer bridges two segments of the
polyelectrolyte.  Internal polarization could also develop at a different time
scale. In order to understand the coupling between chain unfolding dynamics and
interfacial charge relaxation/dissociation dynamics, we study here how $p_x$
and $D_s$ evolves in cycles ($t/\Psi$) of the AC field for different values of
$\Psi$.  $p_x$ is the induced dipole moment $\vec{p}$ of the PE complex in the
field direction. The PE complex, by definition, consists of the chain itself
and the ions condensed inside a tube region of radius $r_t=\lambda_{\rm B}$
surrounding the chain backbone.  $\vec{p}$ is calculated by $\sum_{i} Z_i e
(\vec{r}_i-\vec{r}_{\rm cm})$ where $i$ runs over all the particles in the PE
complex and $\vec{r}_{\rm cm}$ is the center of mass of the chain.  The results
for $|E|=0.3$  are presented in Fig.~\ref{fig:px-Ds_cycle.eps}.

We saw in Fig.~\ref{fig:px-Ds_cycle.eps}(a) that $p_x$ grows exponentially with
time in every half cycle of the square-wave field  and switches its sign
simultaneously with the field.  The growth of $p_x$  saturates to some value if
$\Psi$ is large enough, for example, $\Psi=~10^6 \Delta t$.  It suggests that
there is a relaxation time to polarize the PE complex in an electric field.
Fig.~\ref{fig:px-Ds_cycle.eps}(b) shows that for large $\Psi$, the chain
shrinks right after the field switches the direction, but the chain size does
not reduce down to the zero-field value.  It regains its size in the
corresponding DC field after $p_x$ reaches the saturation value. On the other
hand, for small $\Psi$, the variation of $D_s$ is not significant. To go
further, we drew the parametric plot of $p_x$ and $D_s$ with time in
Fig.~\ref{fig:px-Ds_cycle.eps}(c).  We observed that the trajectory points
travel along a butterfly curve, from the tip of one wing to the tip of the
other wing,  after alternating the field direction, and then wander near the
region of wing tip waiting for the next switch of field direction, as sketched
in the inset of the figure.  The wing bottom occurs at non-zero $p_x$, showing
that the variation of $D_s$ falls behind the $p_x$.  Therefore, the
chain elongation is slaved to the fast dynamics of the interfacial trivalent
counterion dissociation/relaxation time which dominates the polarization.
Moreover, a beautiful shuttlecock pattern is formed when the ensemble of the
data at different field frequencies is looked together in the $p_x$-$D_s$ plot. 

\subsection{E. Polarization time and fluctuation time}
\begin{figure}[htbp]
\begin{center}
\includegraphics[width=\figurewidth,angle=270]{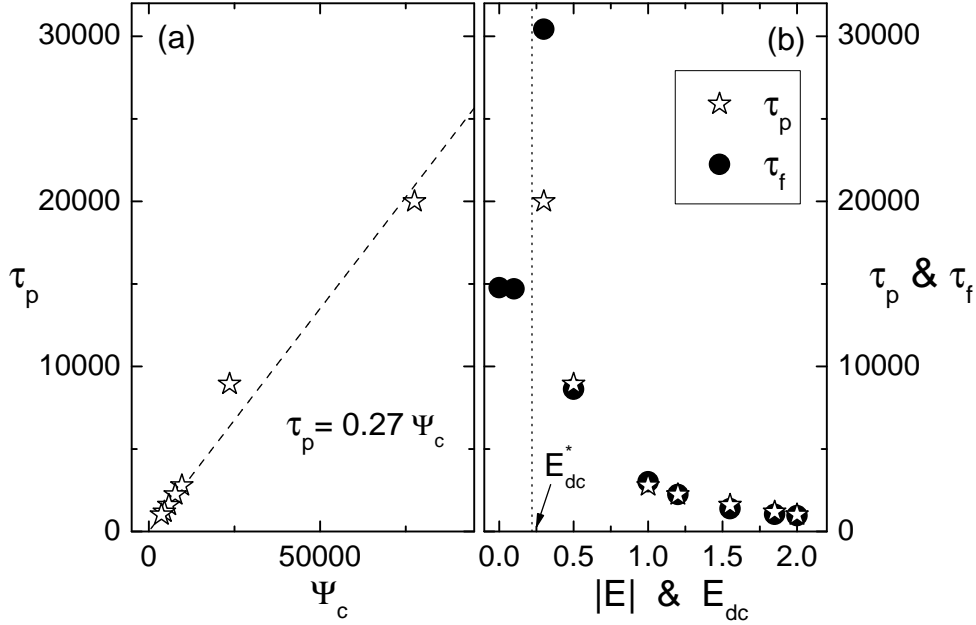} \caption{
(a) polarization time $\tau_p$ vs.~critical period $\Psi_c$, and (b) $\tau_p$
vs.~field strength $|E|$. In panel (b), fluctuation time $\tau_f$ is also
plotted as a function of $E_{dc}$ at the same scale.  The values of $\tau_p$,
$\tau_f$ and $\Psi_c$ are reported in unit $\Delta t$.}
\label{fig:taup_T_E_inflex.eps} \end{center}
\end{figure}

To quantify the findings with our theory, we calculated the polarization time
by fitting exponentially the growth of $p_x$ in a half cycle of field with
large $\Psi$. The mean polarization time $\tau_p$  versus the critical period
$\Psi_c$ to unfold the chain is shown in Fig~\ref{fig:taup_T_E_inflex.eps}(a).

We found that $\tau_p$ depends linearly on $\Psi_c$ with slope equal to
0.27(1). This result interestingly connects the critical points with the
polarization in the supercritical region where the chain is fully unfolded.
The physics can be understood by the following simple picture.  Because the
chain structure is linear in the supercritical region ($\Psi>\Psi_c$), the
polarization time $\tau_p$ can be estimated by the relaxation time $R/v_d$
where $R$ is the relaxation distance of trivalent counterions migrating through
the chain to establish a new polarization when the applied field switches the
direction and $v_d$ is the drifting velocity equal to $\mu_e|E|$.  In our
simulations, $\mu_e \sim 3$ for the trivalent counterions and $|E|\sim 32E_{\rm
dc}^*\omega_c$ for small $\omega_c$ according to Eq.~(\ref{eq:Xi}).  The slope
0.27 yields a relaxation distance of $R \sim 36$ (about $30\%$ of the chain
contour length), which is a reasonable value in view of the profile of charge
distribution along the chain. Fig~\ref{fig:taup_T_E_inflex.eps}(b) shows
$\tau_p$ as a function of the AC field strength $|E|$.  One can see that the
stronger the field strength, the shorter the polarization time. It is because
ions and PEs are dragged at faster speeds when the applied field is stronger.
Moreover, detailed study showed that in the region of strong field
$|E| \gg E_{dc}^*$, $\tau_p$ is inversely proportional to $|E|$.  Also, in the
limit where $|E|$ tends toward $E_{dc}^*$, the product of $\tau_p^2$ and
$|E|^2-E_{\rm dc}^{*2}$ is a constant.  The two behaviors can be deduced easily
from our theory, known that $\Psi_c$ and $\tau_p$ are identical within a
factor. Our simulations are in good agreement with experiments of field-induced
decondensation~\cite{porschke85b}, in which the reciprocal of the relaxation
time of decondensation exhibits a linear relation to the field strength when
the field is strong, and deviates from the relation when the field is weak.

Recently, simulations done by Liu et al.~\cite{liu10} suggested that the
intrinsic chain relaxation frequency is the upper bound of the critical AC
frequency to stretch a PE chain. This bound is quite crude in light of the
current result. To further verify the importance of the
dissociation/condensation of the interfacial trivalent counterions, we
calculated the autocorrelation function of $D_s$, defined by $$
R(t)=\frac{\left<D_s(t)D_s(0)\right>-\left<D_s\right>^2}
{\left<D_s^2\right>-\left<D_s\right>^2},$$ in different DC electric fields.
The correlation time $\tau_f$, calculated by fitting $R(t)$ function
exponentially, represents physically the fluctuation time of chain size under
the action of a DC electric field. The results have been plotted in
Fig~\ref{fig:taup_T_E_inflex.eps}(b) as a function of the field strength
$E_{\rm dc}$ at the same scale.  We found that $\tau_f$ is a constant when
$E_{dc}<E_{dc}^*$.  A large jump occurs at $E_{dc}=E_{dc}^*$ and then, $\tau_f$
decreases inversely with increasing $E_{\rm dc}$. A striking finding is that
$\tau_f$ is basically coincident with $\tau_p$ in the region $E_{dc}>E_{dc}^*$.
We know that the PE chain is in a dynamic equilibrium with the ions surrounding
it. The dissociation or condensation of a counterion away or onto the chain
will locally break the chain stability and cause the fluctuation of the chain
size.  The coincidence between these characteristic times shows the importance
of the ion fluctuations on chain unfolding. The critical frequency to stretch a
chain in a square-wave AC field is mainly determined by the
dissociation/relaxation of the condensed interfacial trivalent counterions,
which takes the value of the fluctuation time of chain size in the counterpart
DC electric field. 
 
\section{IV. Summary}
We have revealed the key conditions to unfold collapsed PE chains in
square-wave AC electric fields: the field strength must be larger than the critical
DC field to stretch the chain and the frequency is smaller than the inverse of
the counterion dissociation/relaxation time under a quasi-DC field. Moreover,
when both conditions are satisfied, there is a strong correlation between the
critical field strength and the field frequency. A simple scaling theory, based
upon Maxwell-Wagner dielectric theory, can quantitatively capture this critical
field-frequency correlation for PE unfolding in AC fields.

\begin{acknowledgments}
This material is based upon work supported by the National Science Council, the
Republic of China, under Grant No.~NSC 97-2112-M-007-007-MY3.  Computing
resources are supported by the National Center for High-performance Computing.
HCC is supported by DTRA (HDTRA1-08-C-0016).
\end{acknowledgments}


\begin{thebibliography}{2}
\bibitem[{\citenamefont{note-epsilon}(2010)}]{note-epsilon} 
\bibinfo{note}{In our system, the unit of conductivity is $e^2/(k_B T\sigma\tau)$ 
which corresponds to 17,200 siemens per meter (S/m).  }
\end{thebibliography}

\end{document}